\documentstyle[preprint,revtex]{aps}
\begin{document}
\draft
\begin{title}
Tuning of the CDW in the Halogen-Bridged
\end{title}
\begin{title}
Transition-Metal Linear-Chain Compounds
\end{title}
\author{M. Alouani and J. W. Wilkins}
\begin{instit}
Department of Physics, The Ohio State University, OH 43210-1368\\
\end{instit}
\author{R. C. Albers and J. M. Wills}
\begin{instit}
Los Alamos National Laboratory, Los Alamos, New Mexico 87545\\
\end{instit}
\receipt{April 1993}
\begin{abstract}
Local-density-approximation calculations are used to show that the
metal-metal distance along the chains controls the charge-density wave
(CDW) in halogen-bridged transition-metal linear-chain (MX) compounds.
The strength of the CDW can be understood in terms of a two band
Su-Schrieffer-Heeger model if a hard-core ion-ion repulsion potential
is also added.  We predict a second-order phase transition from an
insulating to a semi-metallic ground state and explain trends in
dimerization, bond-length ratios, band gaps, and Raman breathing modes
in terms of the metal-metal distance.
\end{abstract}
\narrowtext
During the last few years halogen-bridged transition-metal linear-chain
(MX) compounds have been extensively studied
\cite{kell82,clar84,whangbo,sakai,kuroda,okamoto91,okamo92,int74,dono91,scott92,kanner93,gamm91a,aaw}.
One reason for the increased interest in these materials is that their
various competing ground-states such as charge-density-wave and
spin-density-wave are sensitive to tuning by chemical substitutions,
pressure, or doping
\cite{kell82,clar84,whangbo,sakai,kuroda,okamoto91,okamo92,int74,dono91,scott92,kanner93}.
In particular, progress has been achieved recently and independently by
groups in Japan \cite{sakai,kuroda,okamoto91,okamo92} and in Los Alamos
\cite{dono91,scott92,kanner93} in tuning the charge density wave
(CDW) strength over a wide range by using different counter-ions and
equatorial ligands.

In this Letter, we use local-density-approximation (LDA) full-potential
linear muffin-tin-orbital (FPLMTO) calculations to study the tuning of
the CDW of the MX systems as a function of the metal-metal (M-M)
distance along the chain. By comparing experimental results for MX (X=
Br, Cl) systems with varying M-M distances to our calculated results for
Pt$_2$X$_6$(NH$_3$)$_4$ (X= Br, Cl) under uniaxial stress, we show that
{\it the key parameter} is the M-M distance.  Thus, the effect of the
different ligands on the properties of these systems is indirect; they
tune the M-M distance which is really responsible for the change in the
electronic properties.  Moreover, we also demonstrate that an extended
Su-Schrieffer-Heeger (SSH) model with a universal set of parameters fits
the entire class of Pt charge-density-wave MX materials.  Thus, we have now
provided a crucial theoretical underpinning that explains the CDW MX
systems.

Our principal results for the uniaxial stress are:  (i) The
dimerization, the condensation energy, the Raman breathing mode, and the
band gap are gradually reduced, and vanish when the Pt-Pt distance is
reduced by nearly 7\% from its ground-state value.  (ii) Under the same
conditions, we predict that the PtBr MX system exhibits a second-order
phase transition from an insulator to a semi-metal.  (iii) The two-band
SSH model \cite{ssh} describes the tuning of the CDW strength {\it only
if} we also introduce a repulsive potential between the metal and halogen
ions.

The calculations presented in this paper are based on the LDA with an
all-electron scalar-relativistic FPLMTO basis-set as described in our
previous work \cite{aaw}.  This method was successfully used to  show
that the dimerization and the insulating ground state of
Pt$_2$Br$_6$(NH$_3$)$_4$ are correctly predicted provided that we
include the full ligand structure \cite{aaw}.  Under uniaxial stress, we
rigidly move the ligands with the metal atoms.

A single MX chain (M = Ni, Pd, or Pt; and X = Cl, Br, or I) consists of
alternating MX units, arranged such that two neighboring X atoms move
closer to one of the M atoms \cite{kell82,clar84}.  This dimerization is
schematically represented by
...M$^{3-\delta}$...XM$^{3+\delta}$X...M$^{3-\delta}$...XM$^{3+\delta}$X...
Due to the strong electronegativity of the X atoms, the dimerization
causes an alternation of the valence character of the metal atoms
between $3-\delta$ and $3+\delta$. The strength of the CDW produced is
then directly proportional to the dimerization.

{\it Metal-insulator phase transition.}
Figure \ref{fig1} shows the LDA total energy of the PtBr system for
Pt-Pt distances less than the experimental lattice parameter (5.55 \AA)
as a function of the dimerization $\xi$. Under uniaxial stress, the
double-well structure in the energy collapses to a single well, and the
condensation energy (the difference between the energy at zero
dimerization and the minimum energy) and the dimerization decreases as
the Pt-Pt distance is reduced.  This agrees with earlier work by Sakai
{\it et al.} on Pt(en)Cl(ClO$_4$)$_2$ under pressure \cite{sakai}.  At a
Pt-Pt distance of 5.18 \AA~ (which is equivalent to a uniaxial stress of
89 kbars based on the first derivative of the total energy versus cell
volume) the condensation energy and the dimerization vanish
simultaneously, and the system exhibits a second-order phase transition
from an insulating to a semi-metallic ground-state.  This is confirmed
by the calculated band gap which decreases gradually and closes when the
Pt-Pt distance is near 5.18 \AA~ (see the right side of Fig.
\ref{fig2}).  The possibility of this transition in
[Pt(NH$_3$)$_4$XPt(NH$_3$)$_4$X]$^{4+}$ chains was first suggested by
Whangbo and Foshee using an extended H\"uckel method \cite{whangbo}. It
is also consistent with pressure measurements on the PdBr system, where
140 kbar of hydrostatic pressure caused a nine-order-of-magnitude
increase in electrical conductivity \cite{int74}, due to an increased
carrier concentration (because of the reduced gap). However, beyond 140
kbar of pressure a leveling off of the conductivity versus pressure and
an increase of the activation energy was observed upon application of
pressure up to 350 kbar \cite{int74}. Similar conclusions were reached
by other researchers \cite{kuroda,okamoto91,dono91} for
Pd(chxn)Br(Br)$_2$, Pt(en)Cl(ClO$_4$)$_2$, and Pt(en)Br(ClO$_4$)$_2$
systems respectively.

Following the suggestion of Interrante {\it et al.} \cite{int74} that a
compression of the metal-ligand bond length may increase the band-gap,
we have performed self-consistent calculations of the PtCl band
structure in which the interchain distance is decreased by 5\% without
modification of the M-M distance of 5.45 \AA~ along the chain.  When the
ligand-metal distance is held constant the band gap is unaffected,
whereas a 5\% decrease in the ligand-metal bond-length increases the
energy band gap by 0.25 eV. These results coupled with the decrease in
the band gap when the metal-metal distance along the chain decreases
suggest an {\it intricate} interplay between the ligand-metal distance
and the conduction process along the chain, and may serve as a
qualitative explanation for the increase of the band-gap beyond some
critical pressure. Thus, real pressure is a more complicated situation
than the uniaxial pressure generated by chemical subsitutions, where the
ligand-metal distance is constant, the CDW state is controlled by the
M-M distance, and a metallic state should occur as shown by our
calculation.

{\it Softening of the Raman mode.}
The use of a harmonic oscillator model in the effective potential given
by the total energy curve allows us to estimate the optical (Raman)
breathing-mode phonon frequency as a function of the Pt-Pt distance.
This breathing mode is an oscillation of the two Br chain atoms around a
Pt atom at rest.  Since the curvature in the total energy curve (at its
minimum) decreases under uniaxial pressure, this frequency is expected
to decrease.

Figure \ref{fig2} displays our calculated Raman phonon frequencies and
band gap as a function of the Pt-Pt distance.  We observe an almost
linear softening of the Raman mode as the M-M distance is shortened.  In
the vicinity of the phase-transition, where the curvature of the total
energy becomes flat (cf. Fig. \ref{fig1}) the phonon mode softens
considerably. The phonon frequencies for Pt-Pt distances below 5.37 \AA~
are not displayed, because the use of a simple harmonic oscillator to
calculate the phonons is dubious since the lattice distortion is
comparable to the zero-point-motion, and the phonon frequency is
comparable to the condensation energy.  In this case, thermal effects
and zero-point motion, which may cause a breakdown in the
Born-Oppenheimer approximation, must be taken into account.

Figure \ref{fig3} gives a comparison between the calculated short
Pt$^{3+\delta}$-X and long Pt$^{3-\delta}$-X bond lengths [X = Br (open
circles), Cl (filled circles)] versus the Pt-Pt lattice parameter with
the different experimental results obtained for different MBr/Cl
compounds (M = Pt, Pd, and Ni). The different ligands and counter-ions
produce a range of M-M distances for the different compounds.  For
example, experimentally, substitution of the ClO$_4$ counter-ion by Br
decreases the metal-to-metal distance along the chain because the
hydrogen bond between NH$_2$ of ethylenediamine (en) or
(1R,2R)-cyclohexanediamine (chxn) ligands and Br counter-ion is
strengthened \cite{okamo92}. The substitution of (chxn) ligand by (en)
ligand reduces the metal-metal distance because (chxn) has a non-planar
structure whereas (en) is planar \cite{okamo92}.

The experimental results of all these different materials track the
calculations under uniaxial pressure. This strongly suggests that the
M-M distance is the key parameter controlling the CDW strength.

In addition, the short M$^{3+\delta}$-X bond distance remains
essentially constant throughout the series of compounds presented,
revealing the contraction of the long M$^{3-\delta}$-X bond as
responsible for the large change in the dimerization, therefore reducing
the strength of the CDW.  Our calculations describe the experimental
situation rather well, even though the experimental results are mostly
for different MX chain systems with different ligands. But this
surprising agreement can be understood qualitatively as follows.  In our
previous work, we have shown that the dimerization in PtBr system is
driven by an electron-phonon coupling confined to a single chain
\cite{aaw}.  This strong electron-phonon coupling reduces the short M-X
distance to the sum of the ionic radii of the metal and the bridging
halogen ions \cite{okamo92}.  Thus, the effect of uniaxial stress is to
decrease the long M-X bond length which is much softer than the shorter
one.  To reduce the short bond length, one has to act against the
strongly {\it anharmonic potentials} of the M$^{3+\delta}$ and X ions
\cite{shan69}.

We have analyzed quantitatively the effects of uniaxial stress on the strength
of the CDW in terms of a two-band single-chain SSH model \cite{gamm91a}.
\begin{eqnarray}
\label{eham}
H_0 &=& \sum_{l\sigma}\biggl\{\bigl(-t_0 + \alpha \Delta_{l}\bigr)
  \bigl(c_{l\sigma}^{\dagger} c_{l+1\sigma}^{\vphantom+}
+ h.c.\bigr)
+~\bigl[(-1)^{l}e_{0}-\beta_{l} \bigl(\Delta_{l}+\Delta_{l-1}\bigr)\bigr]
c_{l\sigma}^{\dagger} c_{l\sigma}^{\vphantom+} \biggr\}
\nonumber\\
&&+ \sum_{l,j=0,1} K_{MX}^{(2j)} \Delta_l^{2(1+j)}
+ \sum_{l,j=0,2}K_{MM}^{(j)} (\Delta_{2l} + \Delta_{2l+1})^{2+j}\;,
\end{eqnarray}
\narrowtext
\noindent
where $c^\dagger_{l,\sigma}$ ($c_{l,\sigma}$) creates (annihilates) an
electron at site $x_l$ with spin $\sigma$, and where M and X occupy even
and odd sites, respectively.  The electron-phonon parameters $\alpha$,
$\beta_{2l}=\beta_M$, and $\beta_{2l+1}=\beta_X$ couple the lattice
(through $\Delta_l=x_{l+1}- x_l$) to the electronic structure.  The
energies $\pm e_0$ and the $t_0$ are the M and X energy levels and the
hopping integral at zero dimerization, respectively.  The spring
constants $K_{MX}^{(j)}$ between M and X and $K_{MM}^{(j)}$ between M
and M model represent elastic energies that model the rest of the
electronic structure that is not included explicitly in the
one-dimensional electronic structure \cite{kmx}.


The bands in the vicinity of the Fermi energy deviate slightly from the
nearest-neighbor tight-binding bands of the SSH model, Eq. (1).
Nonetheless, this model, with only {\it one set} of parameters, fits
nicely the relevant bands for any Pt-Pt distance.

{\it Anharmonic potential.}
One failure of the SSH model is its prediction of an almost constant
ratio of the dimerization to the M-M distance (about 3.6\% for PtBr) for
all M-M distances, in disagreement with experiment and LDA (which
predicts a vanishing dimerization for a sufficiently short M-M distance).
We have extended the model by adding a hard-core ion-ion repulsive potential
between the metal and halogen atoms along the chain.  For simplicity, we
have used a next-neighbor repulsive potential.  This changes the model
Hamiltonian to
\begin{eqnarray}
\label{eham1}
H&=& H_0 + \sum_{l}{2A \over (R_{MX}^{(0)}+ \Delta_{l} - R_c)^{q}}\;
\end{eqnarray}
\narrowtext
\noindent
where $A$ and $q$ are constants representing the strength and hardness
of the potential, $R_{MX}^{(0)}$ is the M-X distance at the experimental
equilibrium ground-state at zero dimerization, and $R_c$ is a hard-core
distance between M and X \cite{parameter1}.

The Hamiltonian of Eq.\ (\ref{eham1}) fits the LDA total energy of PtBr
under uniaxial stress rather well (see Fig. \ref{fig1}, dashed curves).
The parameters used to describe the LDA total energies are
\cite{parameters} given in Table \ref{table1}.  In Fig. \ref{fig3} we have
also depicted the short M$^{3+\delta}$-X and the long M$^{3-\delta}$-X
bond lengths predicted by the model given by Eq. (2) as a function of
the Pt-Pt distance.  Notice that the model is in good agreement both
with the LDA and the experimental results.  So, the constant
M$^{3+\delta}$-X short bond is explained as a result of the {\it
competition} between a strong {\it electron-phonon coupling} acting
along the chain and a {\it hard-core repulsive potential} between the
metal and halogen ions.

To further illustrate the model given by Eq.\ (2), we have calculated
the short and long bond for Pt\underbar{Cl}.  The LDA is used to get the
dimerization for the ground state (Pt-Pt distance of 5.27 \AA~
\cite{summa}), then the model is fitted to the LDA results with the
requirement that the short bond-length is equal to the LDA prediction.
This procedure provided the model parameters shown in Table
\ref{table1}. Figure \ref{fig3} shows that our prediction for PtCl
compares favorably with the available experimental results.  The model
also predicts a metallization for PtCl at a Pt-Pt distance of 4.78 \AA.
Notice that the Ni-X sytems (numbers 6 and 13 in the figure) don't
dimerize; in fact, they are magnetic, which opens a gap that prevents
them from being metallic.

In conclusion, we have found that LDA correctly predicts the variation
in CDW strength of PtBr and PtCl under uniaxial stress, and that the key
parameter is the M-M distance.  We have shown that the dimerization, the
band gap, and the breathing phonon mode all decrease linearly with the
M-M distance in agreement with experiments for related systems
\cite{sakai,kuroda,okamoto91,okamo92,int74,dono91,scott92,kanner93}.
At a Pt-Pt distance close to the sum of the ionic radii a second-order
phase transition to a metallic state appears.

Finally, we have shown that the SSH model required the addition of a
hard-core repulsive potential between the metal and the halogen ions to
reproduce the correct tuning of the CDW strength in the MX systems.  This
shows that the constant short M$^{3+\delta}$-X distance is an
equilibrium distance resulting from a {\it competing} strong
electron-phonon coupling and a hard-core repulsive potential between the
metal and the halogen ions.

We thank R. H. Mckenzie for stimulating discussion. Partial support was
provided by the Department of Energy (DOE) - Basic Energy Sciences,
Division of Materials Sciences.  Supercomputer time was provided by the
Ohio Supercomputer and by the DOE.

\figure{\label{fig1}
LDA total energy of PtBr as a function of dimerization ratio $\xi$ (the
ratio of half the difference between the two M-X distances to the M-M
distance in \%) for different Pt-Pt distances. The Pt-Pt distance are,
from the upper curve to the lower one, 5.18 \AA, 5.29 \AA, 5.37 \AA,
5.45 \AA, and 5.55 \AA. The filled circles are the calculated LDA total
energies and the full curves are obtained using a polynomial
least-squares fit of order four. The dashed curves represent the total
energies as obtained from the Hamiltonian given by Eq. (2) (see text).
An energy offset of 0.05 eV instead of the calculated energy offsets has
been added between the different curves for clarity of presentation
\cite{covera}. Under uniaxial stress the size of the dimerization and
the dimerization energy are reduced and, at Pt-Pt distance of {5.18 \AA}
the dimerization vanishes and a second-order phase transition to a
metallic state appears.
}
\figure{\label{fig2}
Calculated (open circles) and measured \cite{kell82} (filled
circle) Raman breathing mode, and
calculated band gap energy (filled triangles) as a function of the Pt-Pt
distance (the experimental band gap under pressure decreases
\cite{kanner93} but the absolute magnitude is not yet known).
The linear softening of the breathing phonon mode is in good
agreement with  the experimetal results of Ref. \cite{sakai,kanner93}
and the results (long dash) of the model given by Eq. (2).}
\figure{\label{fig3}
LDA calculated short M$^{3+\delta}$-X and long M$^{3-\delta}$-X bond
lengths for Pt$_2$Br$_6$(NH$_3$)$_4$ (open circles) and
Pt$_2$Cl$_6$(NH$_3$)$_4$ (filled circles) as a function of M-M distance.
The filled triangle symbols represent experimental data for various MX
systems \cite{okamo92,scott92}: 1-Pt$_2$Br$_6$(NH$_3$)$_4$;
2-[Pt$_2$(en)$_4$Br$_2$] (ClO$_4$)$_4$;
3-[Pd$_2$(en)$_4$Br$_2$] (ClO$_4$)$_4$; 4-[Pt$_2$(chxn)$_4$Br$_2$]
(Br)$_4$; 5-[Pd$_2$(chxn)$_4$Br$_2$] (Br)$_4$; 6-[Ni$_2$(chxn)$_4$Br$_2$]
(Br)$_4$; 7-[Pt$_2$(chxn)$_4$Cl$_2$] (ClO$_4$)$_4$;
8-[Pt$_2$Cl$_2$(NH$_3$)$_8$] (HSO$_4$)$_4$;
9-[Pt$_2$(en)$_4$Cl$_2$] (ClO$_4$)$_4$;
10-[Pt$_2$(tn)$_4$Cl$_2$] (BF$_4$)$_4$;
11-[Pt$_2$(en)$_4$Cl$_2$] (CuO$_4$)$_4$;
12-[Pt$_2$(chxn)$_4$Cl$_2$] (Cl)$_4$;
13-[Ni$_2$(chxn)$_4$Cl$_2$] (Cl)$_4$.
The solid and dashed curves are guides to the eye. The dot-dashed curve
is the results predicted by the model of Eq. 2. Notice that the
calculated and measured short M-X bond lengths are roughly independent
of the M-M distance and that the agreement between the calculations and
the experimental results is good.
}
\widetext
\begin{table}
\caption{Parameters of the model given by Eq. 2 obtained from the fit to
the LDA results. The units of $2e_0$ and $t_0$ is eV, that of $\alpha$ is
eV/\AA, and that of $K^{(j)}$ is eV/\AA$^{j+2}$. The $\beta$ parameters
are found to be zero.}
\begin{tabular}{cccccccccccc}
& $2e_{0}$ & $t_{0}$ & $\alpha$ & $K^{(0)}_{MX}$ &
$K^{(2)}_{MX}$& $K^{(0)}_{MM}$& $K^{(1)}_{MM}$& $K^{(2)}_{MM}$ & A (meV) &
$R_c$
(\AA) & $q$ \\
\tableline
PtBr & 2.3 & 1.5 & 2.4 & 0.007 & 7.0 & 2.2 & -2.6 & -2.0 & 7.0 & 2.0 & 3 \\
PtCl & 2.9 & 1.6 & 2.3 & 0.0 & 1.7 & 3.9 & -5.5 & -16.0 & 9.4 & 1.8 & 3 \\
\end{tabular}
\label{table1}
\end{table}

\begin{references}
\bibitem{covera} The energy offsets between the curves at zero
dimerization, from top to bottom, are 0.32 eV, 0.17 eV, 0.13 eV, and
0.22 eV/(unit cell). The total energy versus the Pt-Pt distance produced
a global minimum at 5.71 \AA~ and a dimerization of 4.2\%. The
experimental Pt-Pt distance and dimerization are is 5.55 \AA~ and
4.8-5.2\%, respectively \cite{kell82,clar84}. In Fig. 1 only total
energies corresponding to Pt-Pt distances below the experimental lattice
parameter are shown.
\bibitem{parameters} The parameters $e_0$, $t_0$, $\alpha$, $\beta_M$,
and $\beta_X$ are kept fixed to the values obtained from the fit to the
LDA band structure under uniaxial stress.
\bibitem{parameter1} The fit is insensitive to the parameter $q$, and
the parameter $R_c$ is directly proportional to the sum of the radii of
the metal and the halogen ions.
\bibitem{kmx} The $K_{MX}^{(2j+1)}$ are zero because the total energy is
even with respect to the dimerization.
\end{references}
\end{document}